\documentstyle[prl,eqsecnum,aps,epsf]{revtex}

\begin{document}

\draft
\twocolumn
\title{Influence of phase space localization on the energy
diffusion in a quantum chaotic billiard} 

\author{D.A. Wisniacki$^{1}$ and E. Vergini$^{2}$} 

\address{$^1$Departamento de F\'{\i}sica ``J.J. Giambiagi'', 
FCEN, UBA, Pabell\'on 1, Ciudad Universitaria, 1428 Buenos Aires,
 Argentina}

\address{$^2$ Departamento de F\'{\i}sica, Comisi\'on Nacional de
Energ\'{\i}a At\'omica. 
 Av. del Libertador 8250, 1429 Buenos Aires, Argentina.}
 
\date{\today}

\maketitle

\begin{abstract}
 
The quantum dynamics of a chaotic billiard with moving boundary 
is considered in this work.
We found a shape parameter Hamiltonian expansion which enables us to obtain
the spectrum of the deformed billiard for deformations so large as the
characteristic wave length. Then, for a specified time dependent shape
variation, the quantum dynamics of a particle inside the billiard is 
integrated directly.
In particular, the dispersion of the energy is studied in the 
Bunimovich stadium billiard with oscillating boundary. 
The results showed that 
the distribution of energy spreads diffusively for the first oscillations
of the boundary (${\langle \Delta^2\! E}\rangle \!=2 \;D \; t$). 
We studied the diffusion contant $D$ as a function of the boundary velocity
and found differences with theoretical predictions based on random
matrix theory. By extracting highly phase space localized structures from
the spectrum, previous differences were reduced significantly. This fact
provides the first numerical evidence of the influence of phase space
localization on the quantum diffusion of a chaotic system. 
\end{abstract}
\centerline{PACS numbers: 05.45.+b, 03.65.Sq, 03.20.+i}
\narrowtext

\section{Introduction}
The quantum dynamics of a classically chaotic system is in the
present a topic of very active interest. Specially, with the great
development in artificially fabricated small devices as quantum dots
\cite{dot},
where quantum manifestation of classical chaos \cite{Gutz} plays 
an important role.

Systems governed by a parameter dependent Hamiltonian 
$H( \ell )$ are excellent models to study quantum manifestations 
of classical chaos. Dynamical localization \cite{cass} and phase space 
localization (scarring phenomena) \cite{ver1,tom} have been investigated
in those systems.
Recently, Wilkinson \cite{wilki1} studied the dispersion of the energy 
in a generic non integrable system when the parameter is time dependent. 
He observed that it spreads diffusively and found asymptotic theoretical
expresions (using random matrix theory) for the diffusion 
constant in the limits of large and small velocities of the parameter. 
Numerical experiments performed in a 
random matrix model were in good agreement with those predictions.
The physical motivation of these studies is the possibility 
of modelling the
quantum dissipation of a finite-size system of non-interacting fermions, 
based on that 
microscopic diffusive behaviour of the energy dispersion\cite{wilki1}. 

In this paper, we treat the dispersion of the energy
in a more realistic system: a two dimensional chaotic billiard
with moving boundary.  
It is not a simple numerical task to solve the time dependent 
Sch\"{o}edinger equation for a 2-D billiard 
with moving boundary \cite{majo}, specially, it is very lengthy from 
the computationally point of view. 
We found a simple shape parameter Hamiltonian expansion which enables
us to obtain the spectrum (up to second order) and wave functions of
the deformed billiard for deformations so large as the characteristic
wave length in the working energy region. Using that expansion and
defining a shape oscillatory motion, the quantum dynamic of a particle
inside the billiard reduces to a system of coupled linear differential
equations which can be integrated directly. On the other hand, this
expansion can be used to obtain efficiently static properties of the
spectrum like curvature distributions or avoided crossing distributions.

We studied the quantum dynamics of the Bunimovich 
stadium billiard with oscillating walls and calculated the diffusion constant
$D$ as a function of the boundary velocity. Differences with
theoretical predictions were found. For small velocities of the boundary,
the system diffuses more than predictions. 
In this regime, the existence 
of bouncing ball
states enables coherent transport of probability,
so the diffusion is enhanced. On the other hand, for large velocities, the system diffuses less  
than predictions. We propose phase space localization 
as the possible mechanism to reduce diffusion in 
this regime. These assertions are supported with numerical simulations 
on the stadium billiard where the most
localized states in the region under study were extracted. In this case,
the system behaves in good agreement with the theory.

The outline of the paper is as follows. To make it self-contained, 
in Sec. II we give a short
introduction to the main
results obtained previously for the diffusion of the energy in
generic chaotic systems. In Sec. III we show how to expand the
Hamiltonian of a 2-D planar billiard in powers of the shape parameter.
In Sec. IV we give a detailed numerical study of the diffusion constant
$D$ as a function of the velocity of the boundary for the stadium billiard. 
Section IV is devoted to final remarks. We include an appendix
with the expresions required to obtain the parameter
Hamiltonian expansion in the stadium billiard.
 
\section{Quantum energy diffusion in chaotic systems}

Let $H(\ell)$ be a parametric Hamiltonian related to a generic 
chaotic system for all $\ell$. The parameter $\ell$ is 
a time dependent function $\ell (t)$.
We restrict to systems with time reversal symmetry where
the statistical properties of the spectrum
are well described by the eigenvalues of ensambles of gaussian orthogonal 
random matrices (GOE) \cite{porter}.
At any $\ell(t)$, the system admits an energy spectrum given by 
the eigenvalue problem for the instantaneous Hamiltonian
\begin{equation}
\hat{H}(\ell(t))\; \psi _{\mu} (\ell(t),{\bf r})=
{k_{\mu}^{2} (\ell(t)) \hbar ^{2}
\over 2 m} \; \psi _{\mu} (\ell(t),{\bf r}).
\label{hem}
\end{equation}  
The system is prepared at $t=0$ in a highly excited eigenstate $\psi _{\nu}
(\ell(0),{\bf r})$ of the Hamiltonian $\hat{H}(\ell(0))$. 
The state for time $t$ can be expressed in the basis of eigenstates 
of the instantaneus Hamiltonian
(eq. \ref{hem}), the so called adiabatic basis,
\[
\Psi ({\bf r},t)= \sum_{\mu}  a_{\mu} (t) \; \psi _{\mu} (\ell(t),{\bf r})\;.
\]
The dispersion of the energy for the considered state is: 
\begin{equation}
\Delta^2 E (t)= \sum_{\mu} |a_{\mu}(t)|^2 (E_{\mu}(\ell(t))-
E(t))^2,
\label{sigma}
\end{equation}
where $E(t)\equiv \langle \Psi|E|\Psi \rangle =
\sum_{\mu} |a_{\mu}(t)|^2 E_{\mu}(\ell(t))$ is the expectation 
value of the energy as a function of $t$. 

Wilkinson \cite{wilki1} observed that the dispersion of the energy exhibit
a diffusive growth when it is averaged over many states, that is 
\[
\langle \Delta^2 E \rangle (t)= 2\; D\; t.
\]
Using random matrix theory for the statistical properties of the spectrum,
he predicted that the diffusion constant $D$ is an universal function of
$\rho$ (the mean energy density), $\hbar$, $\dot{\ell}$ and the mean value
of the off diagonal elements of ${\partial H / \partial \ell}$ 
that he called $\sigma$.
The dimensionless parameter $\;\kappa=\rho ^2 \hbar \dot{\ell} \sigma\;$
is a measure of the adiabaticity of the variation in the
Hamiltonian. He found the following asymptotic form of the diffusion 
constant for small and large $\kappa$: 
\begin{eqnarray}
D = \left\{
\begin{array}{ll}
2^{-5/4}\pi \Gamma (3/4){1 \over \rho^3 \hbar}
 \kappa^{3/2} & \mbox{if $\kappa \ll 1$} \;, \\
\pi {1 \over \rho^3 \hbar} \kappa ^2 & \mbox{if $\kappa \gg 1$ } \;. 
\label{dif}
\end{array}
\right.
\end{eqnarray}

\section{Hamiltonian expansion for deformed billiards}

This section is devoted to obtain a Hamiltonian expansion for a
particle of mass $m$ inside
a shape parameter dependent planar billiard. For variations
$\delta  \equiv \ell\!-\!\ell_{0}$ of the parameter around 
$\ell_{0}$, we propose the following expansion in the basis 
of eigenfunctions at $\ell_{0}$,
\begin{equation}
H_{\mu\nu}({\ell }_{0}+\delta )\simeq (H_{\mu\mu}+\delta^2\;H''_{\mu\mu}
 /2)\;\delta_{\mu\nu}+ \delta  \; H'_{\mu\nu},
\label{exp}
\end{equation}
with $\delta_{\mu\mu}\!=\!1$, and $\delta_{\mu\nu}\!=\!0$ for $\mu\ne\nu$.
This allows us to express the eigenfunctions of the deformed billiard
in terms of the eigenfunctions at $\ell_{0}$.
Then, when the parameter changes as a function of time, the dynamical
evolution of the particle can be integrated
easily using a standard Runge-Kutta method (see Sec. IV).

Let $\zeta$ be a smooth close curve defining a planar billiard. 
We use a curvilinear coordinate system around the boundary 
with $s$ along $\zeta$ and $z$ perpendicular to it at $s$ ($z\!=\!0$ 
on $\zeta$). Consider now that the boundary is deformed and the 
changes are parametrized by 
\begin{equation}
{\bf r} (s,\delta )= {\bf r_{0}} (s)+z(s,\delta )\;{\bf n}\;,
\label{borde}
\end{equation} 
with ${\bf r_{0}} (s)$ the parametric equation for $\zeta$, and 
${\bf n}$ the outward normal unit 
vector to $\zeta$ at ${\bf r}_{0}(s)$ (see Fig. \ref{corde}).
In the appendix we give $z(s,\delta )$ for the stadium billiard.

Let $\{ \phi_{\mu}({\bf r}) \; ;\; \mu=1,2,...N\}$ be the 
eigenfunctions of the billiard defined by $\zeta$
with eigenwave numbers $k_{\mu}$
around $k_{0}$ ($|k_{\mu}-k_{0}|\leq \Delta k\sim {\rm Perimeter/
Area}$). Taking $\;\hbar^{2}/2m\!=\!1$, we immediately see from 
Eq. (\ref{exp}) that $H_{\mu\mu}=k_{\mu}^{2}$.
The eigenfunctions vanish on $\zeta$ and can be
extended  outside the billiard in a smooth way; 
up to second order as follows:
\begin{equation}
\phi_{\mu}({\bf r_{0}} (s)+z\;{\bf n})=z \;{\partial \phi_{\mu} 
\over \partial {\bf n}} 
({\bf r_{0}}(s)) + {\cal O}(z^{3})\;.
\label{fifuera}
\end{equation}
Moreover, to each function $\phi_{\mu}({\bf r})$ we associate 
the scaling function $\;\phi _{\mu}(k {\bf r}/k_{\mu})\;$.
This family of functions depending on the scaling parameter $k$
verifies Helmholtz equation with wave number $k$.

For small variations of the boundary ($\delta \! \ll \! 1$), it is
valid to use perturbation theory to obtain the eigenfunctions and
eigenvalues of the deformed billiard. With those aproximated solutions
we will obtain a Hamiltonian expansion in powers of $\delta$
which allows us to extend the range of deformations to the order 
of the wave lenght ($ 2 \pi/k_{0}$).

Let $\psi_{\mu}({\bf r},\delta)$ be the eigenfunction of the
deformed billiard obtained from $\phi_{\mu}({\bf r})$ by a continuos
variation of the parameter. As we go to obtain the spectrum up to
second order in $\delta$, we express $\psi_{\mu}$ in terms of
$\phi _{\nu}$ up to first order
\[
\psi_{\mu} ({\bf r},\delta)\!=\!\phi_{\mu}\!
\left(\frac{k_{\mu}(\delta)}{k_{\mu}}{\bf r}\right)\!
+\delta\! \sum_{\nu(\neq \mu)} c_{\mu \nu}
\; \phi_{\nu} \!\left(\frac{k_{\mu}(\delta)}{k_{\nu}}{\bf r}\right),
\]
with $\;k_{\mu}(\delta)\;$ the wave number for which 
$\psi_{\mu} ({\bf r},\delta)$ vanishes on the deformed boundary
(Eq. (\ref{borde})). Then, by expanding around $\ell_{0}$ we have 
\begin{eqnarray}
0\!=\!\frac{d\psi_{\mu}}{d\ell} ({\bf r_{0}}(s),0)= &
[k'_{\mu} \; r_{n}/ k_{\mu}+ z'(s) ]\;\frac{\partial \phi_{\mu}}  
{\partial {\bf n}} ({\bf r_{0}}(s)) + \nonumber \\
   & \sum_{\nu(\neq \mu)} c_{\mu \nu}\;
\phi_{\nu} (k_{\mu} {\bf r_{0}}(s)/k_{\nu})\;,
\label{condicion}
\end{eqnarray}
where primes indicate derivation with respect to $\ell$ at
$\ell_{0}$ and $r_{n}\equiv {\bf r.n}$.
In what follows, we omit the argument $({\bf r_{0}}(s))$ of the 
normal derivatives and drop terms including derivatives of second
order or more. In particular, we take 
$\;\phi_{\nu}(k_{\mu} {\bf r_{0}}/k_{\nu})\simeq (k_{\mu}/k_{\nu}-1)\; 
\partial \phi_{\nu}/\partial {\bf n}\;$. Then, multipying 
Eq. (\ref{condicion}) by $\;\partial \phi_{\mu}/\partial {\bf n}\;$
and integrating on $\zeta$, we have
\begin{equation}
H'_{\mu\mu}\equiv(k^{2}_{\mu})'=
-\oint_{{\zeta}}z'(s) \;
(\frac{\partial \phi_{\mu}}{\partial {\bf n}})^2 ds\;.
\label{nondiag}
\end{equation}
To obtain the last expresion it is necessary to use the following
quasiorthogonality relation \cite{ver}
\[
\oint_{{\zeta}} \frac{\partial \phi_{\mu}}{\partial {\bf n}}
\;\frac{\partial \phi_{\nu}}{\partial {\bf n}}\;
\frac{r_{n}\;ds}{2 k_{\mu}  k_{\nu}}
= \delta_{\mu\nu} + \frac{(k_{\mu}-k_{\nu})}{(k_{\mu}+k_{\nu})}
{\cal O}(1) \;.
\label{ort}
\]
Now, multiplying Eq. (\ref{condicion}) by $\;\partial 
\phi_{\nu}/\partial {\bf n}\;$ and integrating on $\zeta$, we obtain
\[
c_{\mu \nu}(k^2_{\mu}-k^2_{\nu})=-\frac{(k_{\mu}+k_{\nu})}{2 k_{\nu}}
\oint_{{\zeta}}z'(s)\; 
\frac{\partial \phi_{\mu}}{\partial {\bf n}}
\frac{\partial \phi_{\nu}}{\partial {\bf n}}ds.
\]
By perturbation theory the last expresion would be $H'_{\mu\nu}$;
however, it is not symmetrical with respect to $\mu$ and $\nu$.
Working to first order in $\delta$ it is sufficient to replace
the factor $(k_{\mu}+k_{\nu})/2k_{\nu}$ by one (we assume that
$|k_{\mu}-k_{\nu}| \ll k_{0}$); but to second
order it is not the case. Consider two states at the same distance
from $k_{\mu}$; that is $k_{\pm}=k_{\mu}\pm \epsilon$. Then, the
factor in question (to first order in $\epsilon$) is
$1\mp \epsilon/2k_{\mu}$. To solve the problem, we define 
the symmetrical factors 
\[ 
A_{\mu\nu}=2-(k_{\mu}+k_{\nu})/2k_{0}\;,
\label{symm}
\]
which give the same asymmetrical contribution to the state $\mu$ from
states $\pm$. On the other hand,  
we stress that for generic non-integrable billiards and for any
non trivial deformation (a dilation is trivial) the interaction
is of long range. So, to have into acount the finite dimension
of the basis we suggest to multiply the elements of
Eq. (\ref{nondiag}) by
the following cut-off:
\[
{\rm Cf}_{\mu\nu}= \exp \left[-2\;(k_{\mu}^{2}-k_{\nu}^{2})^2/(k_{0}
\Delta k)^2 \right].
\]
Finally, non diagonal
Hamiltonian elements acquire the expresion
\[
H'_{\mu\nu}=-{\rm Cf}_{\mu\nu}\;A_{\mu\nu}
\oint_{{\zeta}}z'(s)\; 
\frac{\partial \phi_{\mu}}{\partial {\bf n}}
\frac{\partial \phi_{\nu}}{\partial {\bf n}}ds.
\]

Working as before, to obtain $H''_{\mu\mu}$ we multilply 
equation $\;0\!=\!d^{2}\psi_{\mu}/d \ell^{2} ({\bf r_{0}}(s),0)\;$
by 
$\;\partial \phi_{\mu}/\partial {\bf n}\;$ and integrate it on $\zeta$.
The
result is 
\[
H_{\mu\mu}''=\frac{3 (H'_{\mu\mu})^{2}}{2 H_{\mu\mu}}-
\oint_{{\zeta}} z''(s) 
(\frac{\partial \phi_{\mu}}{\partial {\bf n}})^2 ds\;.
\]
Figure ~\ref{espectro}
compares the approximated spectrum
obtained from eq. (\ref{exp})
with the exact one for the Bunimovich stadium billiard; 
the agreement is excellent. 

\section{Numerical results}

We have consider the desymmetrized stadium billiard 
with radius $r$ and straight line of lenght $a$. The boundary only
depends on the shape parameter $\ell=a/r$ (the area is fixed to the value
$1+\pi/4$). The parameter $\ell$ oscillates 
harmonically around $\ell=1$ with frequency $\omega$ and amplitude 
$\alpha$, that is, 
\[
\ell (t) = 1 +\, \alpha \, \sin (\omega t). 
\]

\begin{figure}
\epsfxsize=15.0cm
\epsfbox[119 149 529 642]{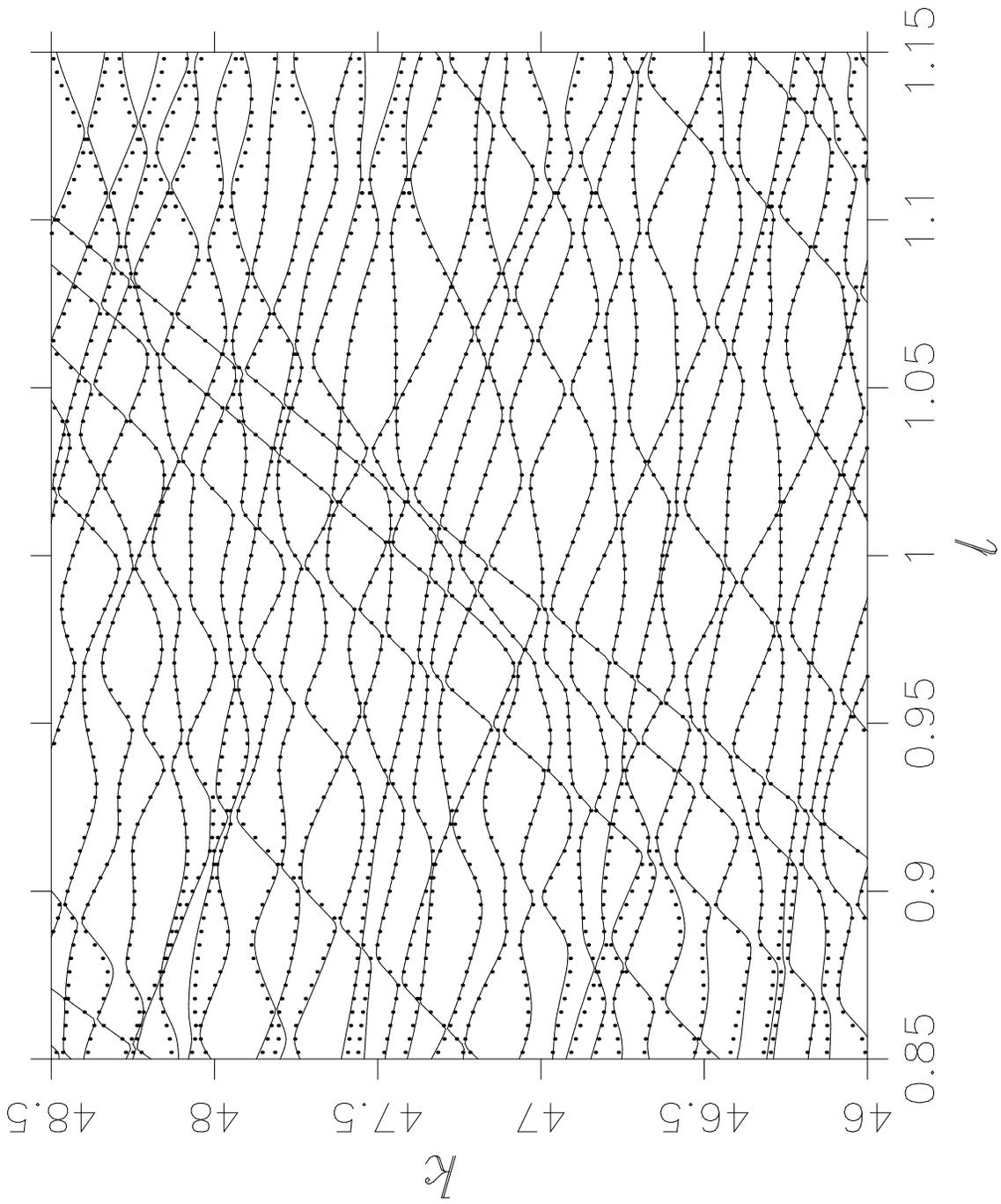}
\vspace{1cm}
\caption{ Approximated spectrum (solid lines) obtained from eq. (3.1)
is compared with the exact one (dots) for the stadium billiar with fixed 
area.}
\label{espectro}
\end{figure}

Sec. II shows universal expresions for $D$ in the asymptotic
limits of $\kappa$; therefore, $\kappa$
is a good quantity to characterize the quantum dynamics of a specific 
chaotic system.
In order to calculate the coefficient $D$, we solved
the time dependent Schroedinger equation for different initial conditions 
\begin{equation}
i \hbar \frac{\partial \Psi(t,{\bf r})}{\partial t} = \hat{H}(\ell(t))
\; \Psi(t,{\bf r}).
\label{sho}
\end{equation}
A very efficient way (computationally) to solve \ref{sho}
is using the parameter expansion of the Hamiltonian obtained in Sec. III. 
The solution $\Psi({\bf r},t)$ is expanded in the basis of 
eigenfunctions at $\ell(0)=1$, 
\begin{equation}
\Psi({\bf r},t)= \sum_{\mu} b_{\mu} (t) \; 
\phi _{\mu} ({\bf r}).
\label{caca3}
\end{equation}   
Although $\phi _{\mu} ({\bf r})$ satifies boundary conditions
for $\ell(0)=1$, it does not cancel outside the boundary 
(see Eq. \ref{fifuera}). 
Of course, the amplitude of the oscillations is limited 
to values of the order of the wave lenght. 
After replacing \ref{caca3} in \ref{sho}, we obtain the system
of differential equations
\[
\dot{b}_{\mu} (t)={-i \over \hbar} \sum_{\nu} H_{\mu \nu}(\ell(t))
\; b_{\nu}\;.
\label{caca4} 
\]
This system was integrated using a standard fourth-order Runge-Kutta method,
and the dispersion of the energy is evaluated using (\ref{sigma})
at times $t=2 \; n \; \pi / \omega$ ($n=0,1,2,...$).
We take ${\hbar}=1$ and $m=1/2$. Then, we used a basis of 103 odd-odd 
eigenfunctions around  $k_{0}=47.3$ at $\ell=1$
(see Fig. ~\ref{espectro}). 
Time evolutions were stopped before the wave functions
had spread into the tails of the spectrum.   

\begin{figure}
\epsfxsize=14.0cm
\epsfbox[12 12 576 687]{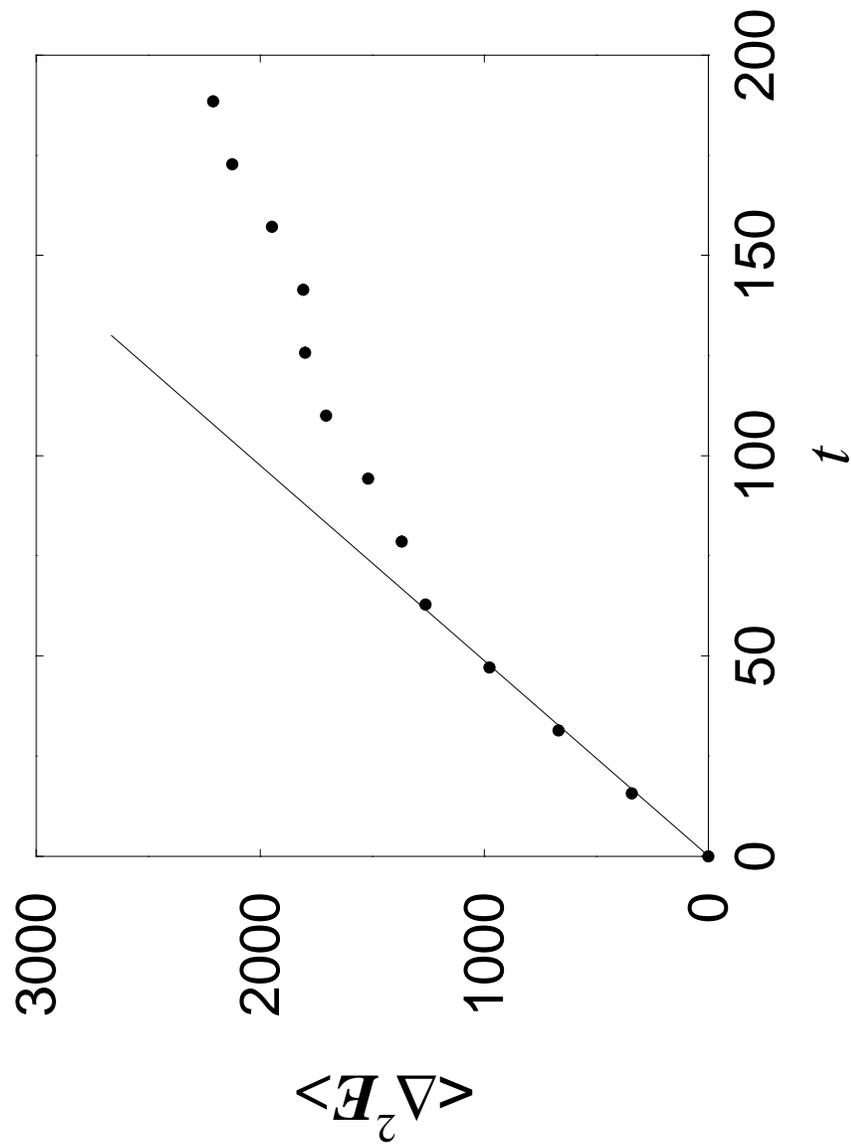}
\caption{Dispersion of the energy $\langle \Delta^{2} E \rangle$ vs $t$
(time) for
$\omega=0.4$. The average is over ten initial states. The solid line
is the best linear fit used to estimate the diffusion constant 
(the slope corresponds to $2 \;D$). }
\label{sigma04}
\end{figure}

As an example, Fig. ~\ref{sigma04} shows the 
dispersion of the energy $<\Delta^{2} E>$ for
$\omega=0.4$ and $\alpha=0.05$. 
The average is over ten initial states. They were
chosen near the center of the spectrum. It is observed clearly a linear
spreading (diffusive behavior) of the dispersion of the energy for the first
oscillations
of the boundary (the slope being equal to $2\;D$); 
later, the spread goes slowly and eventually saturates.

\begin{figure}
\epsfxsize=9.0cm
\epsfbox[0 0 397 631]{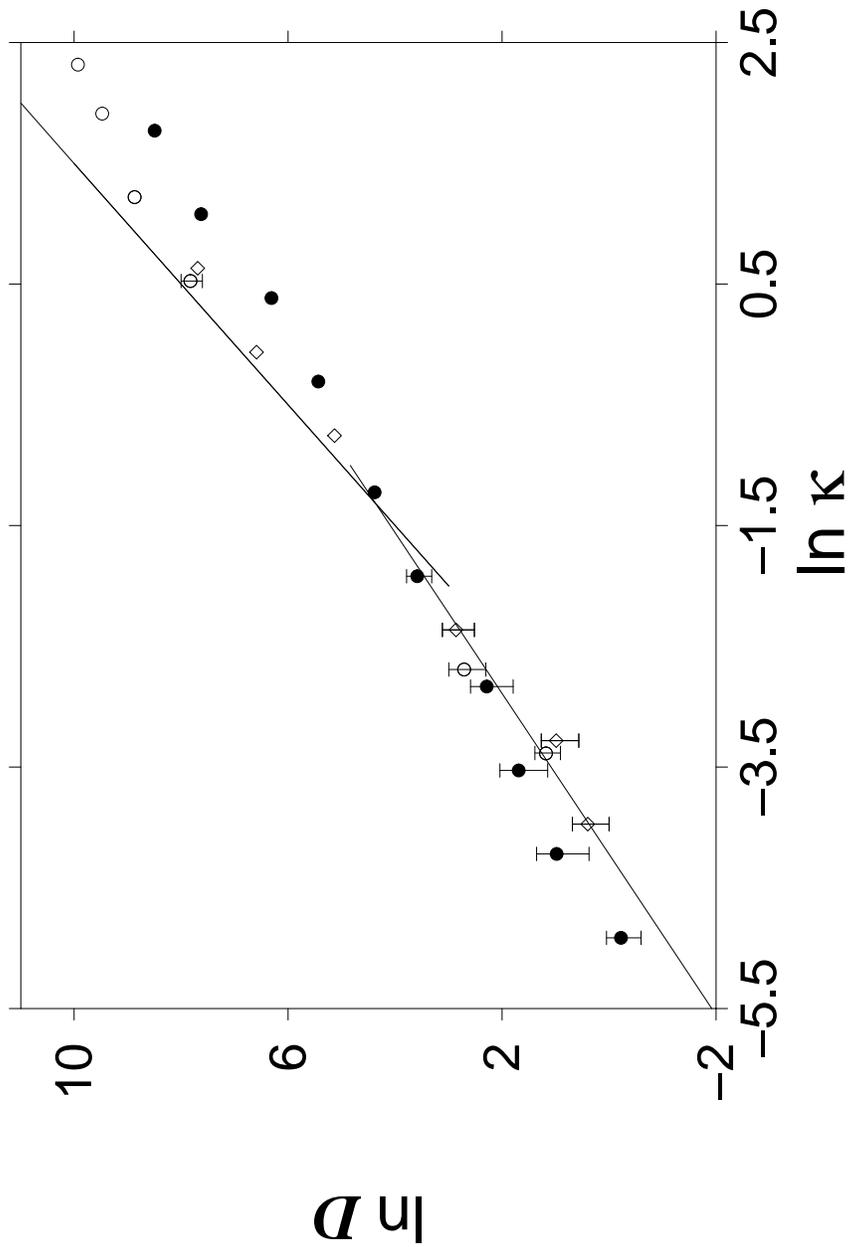}
\vspace{3cm}
\caption{Log-log plot of $D$ vs. $\kappa$. The ($\bullet$) correspond to
 the stadium billiar for
$\omega=0.05,\;0.1,\;0.2,\;0.4,\;1,\;2,\;5,\;10,\;20$ and $40$. 
For the case in which bouncing ball states (see text) were removed,
the ($\diamond$) correspond to $\omega=0.2,\;0.4,\;1,\;5,\;10$ and $20$ 
with $\alpha=0.05$
and, the ($\circ$) correspond to 
$\omega=0.2,\;0.4,\;10,\;20,\;40$ and $60$ with
$\alpha=0.09$.   
It is showed in solid lines the asymptotic theoretical prediccions
for a generic chaotic system with time reversal symmetry (eq. 2.5).
Errors smaler than the symbols are not ploted.}
\label{dvsk}
\end{figure}

Figure ~\ref{dvsk} shows a log-log plot of the difussion constant $D$ 
as a function
of $\kappa$ which is proporsional to the mean
value of the boundary velocity
$<|\dot{\ell}|>=2\alpha \omega/\pi$. Simulations were performed for
$\omega=0.05,\;0.1,\;0.2,\;0.4,\;1,\;2,\;5$, $10$,
$20$ and $40$, with $\alpha=0.05$ (filled circles). 
In all the cases
the average was over ten initial conditions. 
Errorbars were calculated using 
the standard deviation of
the average in the ensamble of initial states. 
The figure shows theoretical predictions in solid straight lines 
for a generic chaotic system with time reversal symmetry (Eq. \ref{dif}),
and we can observe significant differences with the numerical data.

\begin{figure}
\epsfxsize=15.0cm
\epsfbox[119 149 529 642]{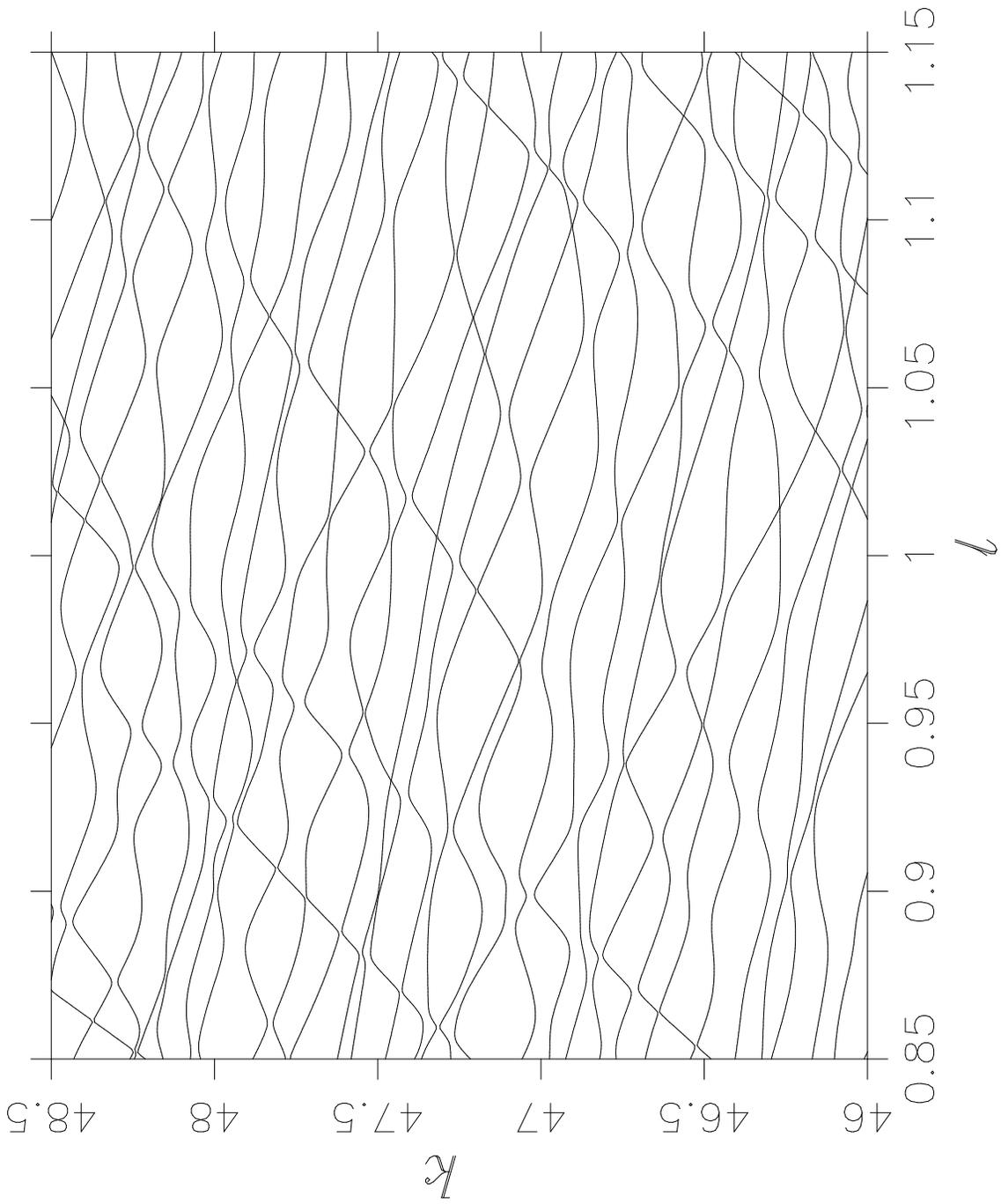}
\vspace{1cm}
\caption{ The same spectrum like in Fig. 2  without the
bouncing ball states with wave numbers  $k=46.4589,\;47.1943,\;47.4220$ 
and $47.7548$ at $\ell=1$.}
\label{sinbb}
\end{figure}

In order to directly see the influence of localized eigenfunctions 
on the quantum dynamics of the system, we extracted from the spectrum
the most localized states. This procedure is very simple by
using the Hamiltonian expansion presented in Sec. III. In order to
extract state $j$,
we remove simply  row and column $j$ from the Hamiltonian matrix.
If at $\ell=1$ this state collide in an avoided crossing it is necessary
to do first the transformation explained in reference \cite{ver1}.
We have removed bouncing ball states with 
$k=46.4589$, $47.1943$,  $47.4220$, and  $47.7548$, 
all of them at $\ell=1$ (see Fig. ~\ref{espectro}). 
The resulting spectrum is showed in Fig. \ref{sinbb}. 
In the same way as before, we have calculated the difusion constant for 
$\omega=0.2,\;0.4,\;1,\;5,\;10$ and $20$  with $\alpha=0.05$
, and for $\omega=0.2,\;0.4,\;10,\;20,\;40$ and $60$
with $\alpha=0.09$. In this case, the results
are very close to theoretical predictions except for
the highest values of $\omega$ (see Fig.~\ref{dvsk}).
 
\section{Final Remarks}
We have studied the spreading of the energy dispersion 
in the Bunimovich stadium billiard with oscillating walls. To the 
best of our knowledge this is the first computation of
diffusion on a quantum chaotic system. We arrived to that 
possibility by using a Hamiltonian expansion which we develop in this
work. 

Numerical simulations show that the energy spreads diffusively 
for a number of oscillations (linear behaviour) and then saturates. 
This saturation phenomenon 
is associated to the fact that the eigenfunctions of the
evolution operator for one period (Floquet states), which 
are localized in energy, inhibit the spread of energy for long times
\cite{gefen,wilki2}.
We calculated the diffusion coefficient (the slope of the initial
linear behaviour) for different velocities of the boundary, observing
significant differences with theoretical predictions based on
random matrix theory. 

In the adiabatic regime (slow boundary velocities),
the system diffuse more than predictions. We associate these differences
to the existence of states strongly localized in phase space (their
Wigner or Hussimi distributions are strongly localized). In particular,
we refer to bouncing ball states which are strongly localized
in momentum space. The interaction between these states and 
generic chaotic ones is much smaller than the typical interaction
between generic states. This fact is reflected on the occurrence 
in the spectrum of 
structures surviving parametric variation; that is, straight lines
interrupted by very small avoided crossings (see Fig. ~\ref{espectro}). Then,
having in mind that in this regime Landau-Zener transitions
\cite{zen,majo} are the mechanism of diffusion, those straight lines
behave like channels allowing a coherent transport of probability
to a wide range of energy in each oscillation. In order to confirm
numerically the above explanation we have extracted from the
spectrum those bouncing ball states living in the region under
study. In this case, the system diffuse according to the theory. 
Certainly, classical billiards with a bouncing 
ball continuous family of orbits display 
anomalous diffusion too \cite{brown}.

As the velocity increases, Landau-Zener probability transitions
across small avoided crossings approach exponentially one. Then,
if the system starts in a bouncing ball state, it moves essentially
up and down by a straight line without mixing. And this situation
mimics the evolution of a regular system where diffusion is 
known to be very small \cite{wilki3}. In the spectrum
without bouncing balls states, observed diffusion agrees with theory
(in Fig. ~\ref{dvsk} the solid line with highest slope) reasonably well.
However, for high velocities simulations and theory fall apart
again showing the beginning of saturation for the diffusion process.

Recently we have proposed \cite{ver1} the elimination of avoided
crossings as the natural mechanism to uncover localized structures
embedded in the eigenfunctions of chaotic Hamiltonian systems.
Precisely, we have shown that many scars of
short periodic orbits are uncover in the stadium billiard transforming
the parametric spectrum in a set of smooth curves which cross among them.
As we mention above, the elimination of avoided crossings can be carried 
out dynamically by increasing the boundary velocity; that is, when
the probability transition is practically one. In the light of
this, saturation of diffusion is expected in chaotic systems for 
high velocities as a consequence of these localized structures
associated to short periodic orbits.

In conclusion, we have verified that asymptotic expresions (\ref{dif})
for the energy diffusion in a chaotic system work very well in a
wide range of boundary velocities except for high velocities where
diffusion saturates. Moreover, if the system includes a fraction
of localized states like bouncing ball (or states living in
a classically regular region which are connected to chaotic states
by tunnelling), diffusion will be anomalous in the adiabatic regime.

\section*{ACKNOWLEDGMENTS}
This work was partially supported by UBACYT (TW35), PICT97 03-00050-01015
and SECYT-ECOS. We would like to thank A. Fendrik, C. Lewenkopf and 
M. Saraceno for useful discussions. 

\appendix

\section{}  
\begin{figure}
\epsfxsize=10.0cm
\epsfbox[170 173 441 618]{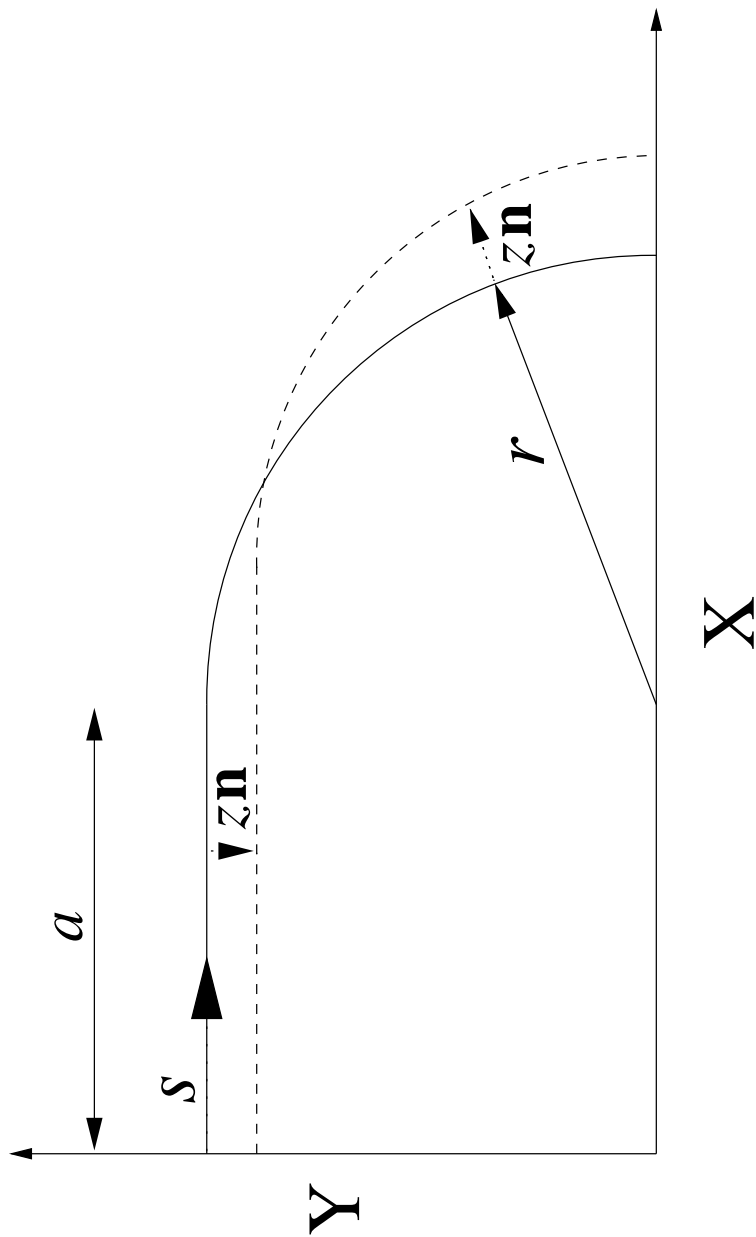}
\vspace{1cm}
\caption{Schematic figure showing the curvilinear coordinate system on the
boundary of the stadium billiard. In dashed line it is shown a deformation
of the billiard. The variations of the boundary are described
by the function $z(s,\delta )$.}
\label{corde}
\end{figure}

In this appendix we show the parametrization used to
describe the deformation of the desymmetrized stadium billiard.
The area is fixed to the
value $A = 1+\pi/4$, so the boundary only
depends on the shape parameter $\ell=a/r$. A point 
${\bf r_{0}}=(x,y)$ 
on the boundary at $\ell_{0}=1$ is given in terms of the 
curvilinear coordinate system defined in Sec. II 
(see Fig. ~\ref{corde})
\,
\[
x(s)= \left\{ \begin{array}{ll}
s & \mbox{if $s \le 1$} \;, \\
1+\sin(s-1) & \mbox{if $s > 1$ } \;. 
\end{array}
\right.
\]  
\[
y(s)= \left\{
\begin{array}{ll}
1 & \mbox{if $s \le 1$} \;, \\
\cos(s-1) & \mbox{if $s > 1$ } \;. 
\end{array} \right.
\]
Then, if the deformed boundaries are described by Eq. (\ref{borde}),  
it is a geometrical problem to show that
$z(s,\delta)\simeq \delta\; z'(s)+\delta^2 z''(s)/2$, with
\[
z'(s)= \left\{
\begin{array}{ll}
- \frac{1}{2 \; A}  & \mbox{if $s \le 1$} \;, \\
\\
(1-\frac{1}{2A}) \sin(s-1)-\frac{1}{2 A} & \mbox{if $s > 1$ } \;. 
\end{array}
\right.
\]
and  
\[
z''(s)= \left\{
\begin{array}{ll}
\frac{3}{4 \; A^2} & \mbox{if $s \le 1$} \;, \\
\\
\\
 \frac{1}{A}-1+(1-\frac{1}{2 A})^2 \sin^2 (s-1)+\;\\
\\
 \frac{1}{2 \; A^2}
-\frac{1}{A} (1-\frac{3}{4 A}) \; \sin (s-1) & \mbox{if $s > 1$ } \;.
\end{array}
\right.
\]

\onecolumn
\clearpage

\clearpage

\clearpage

\clearpage

\clearpage
\end{document}